\documentclass[preprint,review,12pt]{elsarticle}

\usepackage{amssymb}
\usepackage{amsmath,amssymb,lmodern}
\usepackage{mathtools}
\newcommand{\w}{\mathbf}
\newcommand{\de}{\partial}
\newcommand{\e}{\varepsilon}
\newcommand{\epst}{\boldsymbol{\varepsilon}}
\newcommand{\D}{\mathbf{D}}
\newcommand{\E}{\mathbf{E}}

\usepackage{lineno}

\biboptions{sort,compress}

\journal{}

\begin{document}
\begin{frontmatter}



\title{Comprehensive review on Fr\"{o}hlich entropy estimation: a technique to investigate the state of order in condensed matter}
%


\author[Fir]{Jacopo Parravicini}
\address[Fir]{Dipartimento di Fisica e Astronomia, Universit\`{a} di Firenze, IT-50019 Sesto Fiorentino, Italy}\ead{jacopo.parravicini@unifi.it}
\author[Pav]{Gianbattista Parravicini}
\address[Pav]{Dipartimento di Fisica, Universit\`{a} di Pavia, IT-27100 Pavia, Italy}

\begin{abstract}
The so-called \emph{Fr\"{o}hlich entropy} is the entropy variation of a material under the application of an electric field. This quantity can be calculated, under suitable hypotheses, directly from the measured real part of the dielectric function. Although Fr\"{o}hlich entropy is defined for a biased sample, a straightforward physical interpretation correlates it to the state of order of the considered physical system in absence of field. When Fr\"{o}hlich entropy is calculated from experimental results, its trend is able to give several information about the evolution in temperature of the explored compound, especially of its phase transition features. We here provide a comprehensive review of the physical systems (dipolar liquids and nematicons, organic molecular crystals, metallic nanoparticles, inorganic disordered ferroelectrics, etc.) where this approach has been exploited with the aim of evaluating their state of order. The variety of compounds where this method has been applied demonstrates that the estimation of the \emph{Fr\"{o}hlich entropy} can be considered a trustworthy tool for carrying out study on the state of order of different classes of materials. Indeed Fr\"{o}hlich entropy evaluation can be considered a fruitful and reliable investigation technique which can be exploited alongside more usual experimental approaches.
\end{abstract}


\end{frontmatter}
%
%
\section{Introduction and landscape}\label{INTRO}
The issue of investigating the order/disorder degree of a given physical system has addressed by exploiting several experimental methods. The choice of which method should be employed depends on different features, such as what kind of system is considered, which sensitivity is requested, which temporal and/or temperature range should be detected, what complexity and which cost of the used instruments can be faced, etc. The exploited techniques are quite complicated, being employed to investigate both mesoscopic and microscopic realms. A particular attention is devoted to phase transitions in crystalline systems, where several concepts are involved, such as structural symmetries, stability, critical temperatures and amplitudes of phenomena, anisotropies, etc. Morphology and ordering are usually studied by the well known techniques employed in crystallography, such as X-ray diffractometry \cite{Comes1968,Comes1970,Hirota2006}, X-ray absorption spectroscopy \cite{Bunker2010}, EXAFS \cite{Faraci2004,Stohr1991}, electronic diffraction \cite{Tsuda2012,Tsuda2013}, neutron diffractometry \cite{Orauttapong2001,Xu2004}, nuclear magnetic resonance \cite{Hirota2006,Zalar2003,Zalar2005}. These techniques, when resolved in temperature, provide a description of the arrangement of sample internal symmetry, and the relative evolution when macroscopic thermodynamic parameters are changed. Actually they are mainly based on the investigation in reciprocal space, so providing essential information on periodicity, where the presence of a specific kind of crystalline (or not) structure should be inspected. On the other hand, calorimetric techniques are based on different paradigms, which investigate the macroscopic energy states \cite{Ishai2004}. 
%
%
%
%
%

\begin{figure}
	\begin{center}
		\includegraphics[width=0.99\columnwidth]{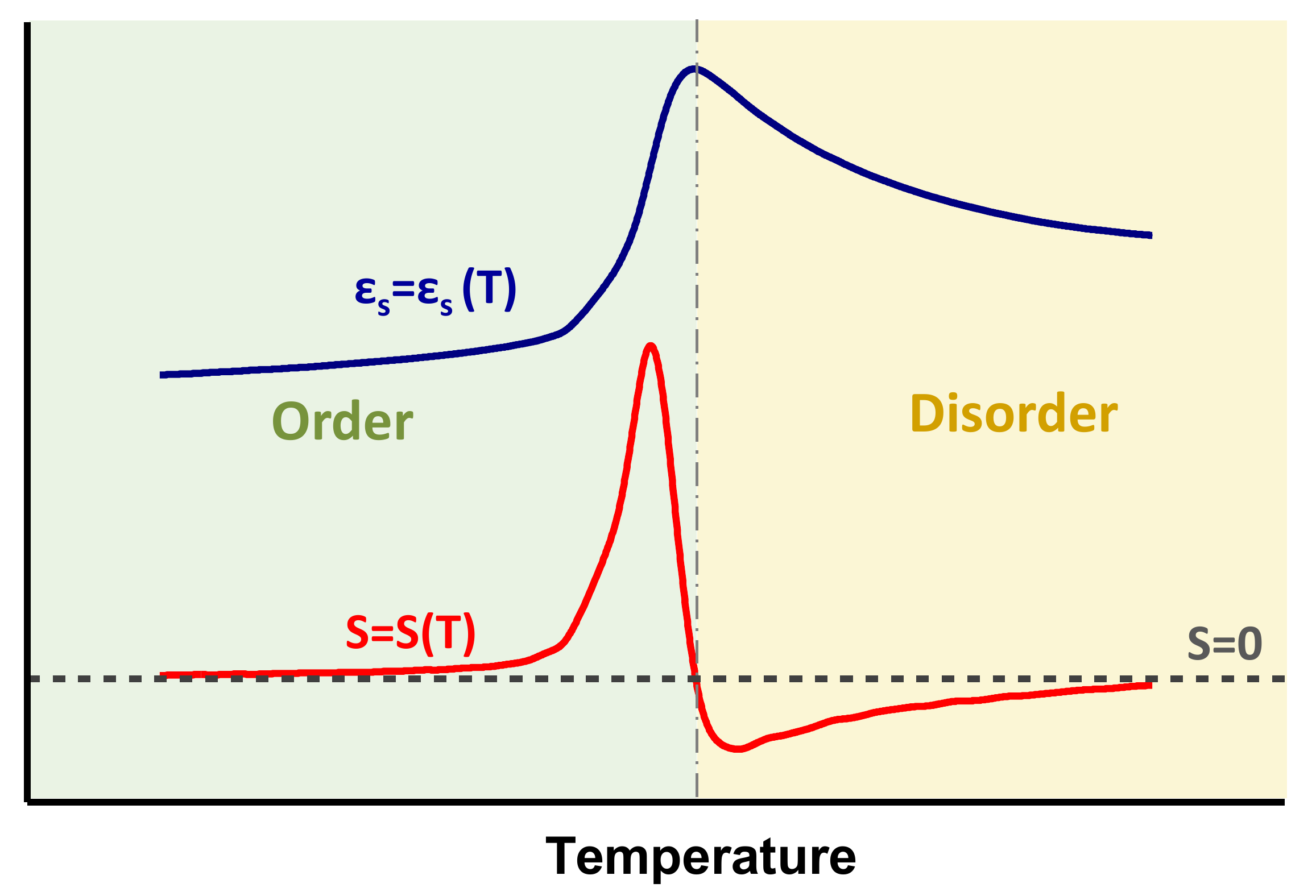}
		\caption{Schematic temperature-dependence of the dielectric constant $\e_{s}$ (blue curve) and of the related entropy change as defined by Eq. (\ref{RelS_Fr}): $S_E\propto \partial\e_{s}/\partial T$ (red curve) due to the field-induced polarization. Fr\"{o}hlich stated that if $S_E>0$ the field creates disorder, if $S_E<0$ it creates order.}\label{SchemaFrohlich}
	\end{center}
\end{figure}

Among experimental techniques, on the other hand, dielectric spectroscopy is typically widely employed to characterize a given material behavior. When the dielectric properties are investigated as a function of temperature, such a technique is able to provide detailed information on the phases of the material. Nevertheless, a dielectric macroscopic technique usually is not considered a method to explore the ordering of a system: in fact a description of the order state of a material would seem to be out of its possibilities. An ideal pathway to obtain order information from dielectric physical quantities would be finding a correlation between them and thermodynamic quantities. The issue of bridging dielectrics and thermodynamics was faced for the first time in the context of the electrocaloric effects \cite{Devonshire1949,Scott2011}; moreover, pyroelectric and piezocaloric effects are also described by equations involving thermodynamic variables \cite{Merz1953,Newhman2005,Nye1985,Ray1982,Narayan2005,Kao2004,Lines2001}. In particular, finding the exact mathematical relationships among thermodynamic variables, electric and magnetic fields, susceptivity and susceptibility, dielectric displacement and magnetization fields, stresses and strains, made possible to extend Maxwell thermodynamic relations for anisotropic compounds \cite{Newhman2005,Nye1985,Devonshire1954,Landau,Tsao1993,Juretschke1977}. However, the first comprehensive thermodynamic theory of a generic dielectrics was developed within a theoretical ``rigorous treatment'' \cite{Johari2016}\footnote{Pg. 6, note 7} by H. Fr\"{o}hlich in Forties \cite{Frohlich1949,Frohlich1958}. The following treatises, whose the main ones are by R. Becker \cite{Becker1964}, L. Landau \cite{Landau}, V. Daniel \cite{Daniel1967}, C. B\"{o}ttcher \cite{Boettcher1973}, and B.K.P. Scaife \cite{Scaife1989,Scaife1998} employ the same concepts and obtain the same relationships. Specifically, Fr\"{o}hlich considered an ideal dielectric undergone to an electric field. He was able to provide equations correlating the induced internal total energy, the entropy variation, the Helmholtz free energy with the dielectric quantities of the given material; they are called, respectively, \emph{Fr\"{o}hlich total energy}, \emph{Fr\"{o}hlich entropy} (FE), and \emph{Fr\"{o}hlich free energy} \cite{Parravicini2016,Parravicini2018,Tan2019}. The most important result of these, however, is the estimation of the entropy induced by the application of a static electric field. Actually the found equation indicates that, under specific conditions, the induced entropy variation is proportional to the derivative of (the real part of) the dielectric function with respect to the temperature.
The deep physical meaning of FE goes far beyond the thermodynamic description of a system in an \emph{under-an-electric-field} condition. We quote Fr\"{o}hlich himself from \cite{Frohlich1958}\footnote{Chapter 1, pages. 12-13}.
\begin{quote}\label{citazione1}
	``Finally, equation 3.12 [$S=S_0(T)+(\de\e_s/\de T) (E^2/8\pi)$, i.e. Eq. (\ref{SE1}) below] shows that the entropy is increased by the field if  $\de \e_s/\de T$ is positive, and decreased if this quantity is negative. Since the entropy is a measure of the molecular disorder, an external field creates order in dipolar liquids and gases for which $\e_s$ decreases with increasing $T$. This may be expected because the field will orientate some of \emph{the dipoles which in the absence of a field are at random}. In some dipolar solids, on the other hand $\e_s$ increases with $T$, which means that an external field increases the disorder. \emph{This too is understandable if one assumes that in the absence of a field the dipoles are in a well-ordered state as may be expected in solids}. The field by turning some of the dipoles into different directions can thus only decrease the existing order.''\footnote{emphasis added}
\end{quote}
Immediately after these sentences, in the same textbook the picture reported in Fig. \ref{SchemaFrohlich} is inserted with the following caption:
\begin{quote}
	``Schematic temperature-dependence of the dielectric constant $\e_s$ and of the entropy change $S\propto \de\e_s/\de T$ due to the polarization by a field. If $S>0$ the field creates disorder, if $S<0$ it creates order. Near the absolute zero of temperature the substance is \emph{already} perfectly ordered. Hence $\de\e_s/\de T$ cannot be negative near $T=0$.''\footnote{emphasis added}
\end{quote}
So Fr\"{o}hlich states that, by analyzing the trend of the field-induced entropy (\emph{Fr\"{o}hlich entropy}), we can reconstruct the arrangement of the system ``in the absence of field''. This fine argument of Fr\"{o}hlich is the foundation of the here discussed experimental approach. On one hand the pure numerical value of FE is only able to provide the thermodynamic quantity when the electric field is on; on the other hand, the analysis of its behavior as a function of the temperature is able to give deep information about the state of the system also in absence of an applied field. Exploiting FE evaluation to study the state of order of a system lays on this clever Fr\"{o}hlich's interpretation.
%
%
%
%

The recalled Fr\"{o}hlich statements indicate that information about the state of order of a given system could be extracted, in principle, from its dielectric response. This gives the possibility of obtaining some thermodynamic information by exploiting dielectric techniques. Moreover, it enables to investigate the state of order through an approach which is usually not considered for such a kind of investigation and which is generally simpler and cheaper with respect to those above-mentioned.

Despite the significance of this thermodynamic approach was recognized by founders of the physics of dielectrics \cite{Landau,Becker1964,Daniel1967,Scaife1998}, no experimental studies involving his theory have been accomplished for more than forty years after Fr\"{o}hlich's publications. The first experimental use of Fr\"{o}hlich equations and interpretation were carried out in 2003 by Gianbattista Parravicini et al. for the investigation of melting processes in confined metallic nanoparticles \cite{Parravicini2003APL,Parravicini2003EPJD,Parravicini2003PSSB}. Subsequently, Fr\"{o}hlich arguments were fruitfully applied on the study of nematic compounds \cite{Jadzyn2007,Jadzyn2008_01,Jadzyn2008_03,Jadzyn2009,Bauman2010}, molecular crystals \cite{Sassella2011,Parravicini2012}, ferroelectric and glassy perovskites \cite{Parravicini2016,Tan2019,LoPresti2020,DelRe2015,Parravicini2017,Huang2019}, polar glasses \cite{Matyushov2016}, polar and glass-forming liquids \cite{Johari2013,Johari2013_02,Johari2016,Samanta2015,Samanta2016}, polymeric films \cite{Kochervinskii2016}, enzymes \cite{Kurzweil2016}. The use of Fr\"{o}hlich's theory in describing so different systems demonstrates that employing its approach represents a powerful technique which, through the description of the order state, may give specific information on the temperature evolution of mesoscopic and microscopic physical systems. This provides results which are congruent and complementary with those obtained by the above-mentioned best-known methods.

In this review we intend to provide up-to-date landscape of the use of Fr\"{o}hlich entropy in experimental studies. We analyze the required theoretical and experimental conditions where it can be used and how it has been exploited for describing several different physical systems, especially about phase transitions. The obtained wide outlook will demonstrate that Fr\"{o}hlich entropy measurements can be accepted in the set of the trustworthy and well-consolidated methods, a further option which can be chosen, in alternative with respect to the standard investigation ways. This approach could be called \emph{Fr\"{o}hlich entropy evaluation} or \emph{Fr\"{o}hlich entropy technique}.

\section{Theory of Fr\"{o}hlich entropy}
The study of entropy in condensed matter pointed out that different entropic contributions can be inspected in materials. The discussion of these is faced in many works (e.g. \cite{Johari2013,Johari2016,Samanta2015,Matyushov2016,Vilar2014,Leuzzi2007,Dyre2009,Johari2000_1,Johari2000_2,Sutton2020}) which involve many subtle features. Instead, we here stress that Fr\"{o}hlich's arguments are solely based on standard principles of thermodynamics and basic electric field relationships. This simplicity assures the general validity of Fr\"{o}hlich relationships and interpretations, which can be assumed as starting point for more detailed topics\footnote{It is worth quoting G.P. Johari that, in 2016, wrote ``In his rigorous treatment of thermodynamic effects of an electric field, Fr\"{o}hlich derived Eq. (1) [the same of our Eq. (\ref{S_E})] for the entropy change by making two assumptions'', from \cite{Johari2016}, pg. 6, note 7.}.

\subsection{Thermodynamic potentials of dielectrics: seminal relationships}\label{SezioneBasic}
Fr\"{o}hlich equations give the correlation between the electric field $\E$, dielectric displacement field $\D$, static dielectric function $\varepsilon_s$, and the total thermodynamic energy $U$, entropy $S$, Helmholtz free energy $F$. Specific hypotheses are assumed to obtain such relationships. First, Fr\"{o}hlich assumes that the considered material does not change in volume so, in particular, electrostriction effects can be neglected \cite{Frohlich1958,Johari2016}. Then, when a reversible isothermal transformation is considered, the variation of these thermodynamic quantities can be written as%
\begin{subequations}
	\label{RelGenerali}
	\begin{align}
		U(T,\E) & =U_0(T)+U_E(T,\E),\label{RelU_I}\\
		S(T,\E) & =S_0(T)+S_E(T,\E),\label{RelS_I}\\
		F(T,\E) & =F_0(T)+F_E(T,\E),\label{RelF_I}
	\end{align}
\end{subequations}%
where $U_0$, $S_0$, $F_0$ are the change of respectively thermodynamic total energy, entropy, Helmholtz free energy in absence of field, while $U_E$, $S_E$, $F_E$ are their respective field-induced variations. The constraint of handling reversible transformations requires that, at each considered temperature, the involved physical variables do not depend on the time \cite{Callen}; this implies that relationships (\ref{RelGenerali}) should hold for static fields only. In his treatise Fr\"{o}hlich starts from the electromagnetic energy $U_e$ correlated with the application of $\E$ to an ideal loss-free dielectric medium. Therefore $U_e$, for unit of volume, is given by%
\begin{equation}\label{Densita}
	U_e= \frac{1}{2}\E\cdot\D.
\end{equation}%
When $\D$ changes with the temperature $T$, from Eq. (\ref{Densita}) we can obtain the energy difference (per unit volume) of the considered dielectric, although such relationship apparently does not contain thermodynamic quantities. If the variation of $\D$ is due to some thermodynamic process occurring to the dielectric, the effective energy provided by the above expression will depend on the feature of the occurring transformation. Becker stressed that for an adiabatic process the variation of electric energy $dU_e$ coincides with the variation of the total thermodynamic energy $dU$ \cite{Becker1964}; on the other hand, when an isothermal process is considered, $dU_e$ coincides with the variation of the Helmholtz free energy $dF$ \cite{Frohlich1958,Becker1964}. We note that in this theory the volume variation of the system is considered negligible, so the changes of Helmholtz and Gibbs free energies coincide. Actually, adiabatic experimental conditions are usually difficult to be treated \cite{Becker1964}. Instead, isothermal situations can be much more easily theoretically treated and discussed, so the Fr\"{o}hlich theory is developed in these conditions. In his treatise, Fr\"{o}hlich considers a system whose volume is maintained constant and the sole independent variables are the temperature $T$ and the external electric field $\w{E}$. We summarize the conditions assumed by Fr\"{o}hlich in four \emph{assumptions about the} features of the \emph{dielectric}, i.e.
\begin{enumerate}\label{F_assump}
	\item linear dependence of $\D$ on $\E$,%
	\item scalar approximation (isotropic dielectric),
	\item loss-free medium (ideal dielectric),
	\item volume variation negligible (no electrostrictions, etc.),
\end{enumerate}
and two \emph{assumptions about the thermodynamics} of the considered process, i.e.
\begin{enumerate}
	\item reversibility,%
	\item isothermal conditions.%
\end{enumerate}
We here report the demonstration given by Fr\"{o}hlich in his original texts \cite{Frohlich1949,Frohlich1958}. All quantities in his calculations were assumed for unit of volume, so we will in this exposition. We start from the assumed linear dependence $D=\e E$\footnote{we here employ the absolute dielectric function $\e\equiv\e_0\e_s$, where $\e_0$ is the vacuum dielectric permittivity and $\e_s$ is the static relative permittivity}, the scalar approximation ($\e$ is a scalar quantity) and the loss-free assumption (i.e. $\e$ is purely real). Moreover, no terms of volume variations will be considered. Fr\"{o}hlich argument starts from the standard relationship which provides the density of electric energy $u_e$ for unit volume of a dielectric material
\begin{equation}\label{Fond}
	u_e=\frac{1}{2}ED.
\end{equation}
When a static field is considered, if the correlation between $E$ and $D$ is linear,
\begin{equation}\label{lineareDE}
	D=\e_0\e_s E,
\end{equation}
we have%
\begin{equation}\label{Densita2}
	u_e= \frac{1}{2}\e_0\e_s E^2,
\end{equation}
where $\e_0$ is the static vacuum constant and $\e_s$ is the relative static dielectric constant, as previously hinted. Fr\"{o}hlich recalls that, when $\e_s$ changes with the temperature $T$, Eq. (\ref{Densita2}) should provide the energy difference (per unit volume) of the considered dielectric. From Eq. (\ref{Fond}), we consider the quantity
\begin{equation}\label{du_e}
	du_e=\frac{1}{2}E dD,
\end{equation}
being the energy variation of the dielectric per unit volume if $D$ is infinitesimally varied. If the variation of $D$ is due to some thermodynamic transformation occurring to the considered dielectric, the effective energy provided by the expressions (\ref{Fond}) and (\ref{Densita2}) will depend on the features of the considered transformation. To discuss this, Fr\"{o}hlich takes into account the two fundamental laws of thermodynamics: for an ideal system of volume $v$ and pressure $p$, the first and second law are respectively given, for a reversible transformation, by%
\begin{equation}\label{leggi}
	dU=\delta Q-p\delta v\;\;\text{and}\;\; dS=\frac{dQ}{T},
\end{equation}
\begin{gather}
	dU=dQ-p dv \label{leggi1}, \\
	dS=\frac{dQ}{T} \label{leggi2},
\end{gather}%
where $Q$ is the heat exchanged by the system, $U$ is the exchanged energy, $S$ is the entropy of the considered system, all calculated for unit of volume. Then, we define the Helmholtz free energy $F$ as
\begin{equation}\label{FHelm}
	F=U-TS,
\end{equation}
which represents the maximum amount of work which can be extracted from a system in an isothermal process\footnote{We recall that Fr\"{o}hlich considers a system whose volume is maintained constant and the sole varying parameters are the temperature $T$ and the electric field $E$}. Eqs. (\ref{du_e}) and (\ref{leggi1}) provide the energy variation $dU$, per unit volume, during a process where the field or the temperature, or both, are slightly varied:
\begin{equation}\label{dU}
	dU=dQ+\frac{1}{2}EdD.
\end{equation}
Such relationship has a structure similar to that of the state equation (\ref{leggi1}) where pressure and volume are replaced by $-E$ and $D$ respectively. Nevertheless, in Eq. (\ref{dU}) the dependence of $D$ on temperature must be taken into account. When Eq. (\ref{lineareDE}) holds, we can write
\begin{equation*}
	dD=d(\e_0 \e_s E)=\e_0\left(\e_s dE+ E \frac{\partial \e_s}{\partial T}dT \right),
\end{equation*}
which means that a variation of $D$ may be written as the sum of two contributions: a change in the field-strength $E$ at constant temperature, and a change in temperature at constant $E$. Here it is useful to employ $T$ and $E^2$ as the independent variables. So we can write the first law of thermodynamics given by Eq. (\ref{dU}):
\begin{gather}
	dU=dQ+\frac{1}{2}\e_0\e_s d(E^2) +E^2 \e_0 \frac{\partial\e_s}{\partial T}dT= \nonumber \\
	=\frac{\partial U}{\partial (E^2)}d(E^2) +\frac{\partial U}{\partial T}dT. \label{dU_ext} \\
\end{gather}
A further relation will now be obtained from the entropy law according to which $dS$, given by Eq. (\ref{leggi2}), is a total differential. This means that a unique function $S=S(T, E^2)$ must exist such that
\begin{equation}\label{dS1}
	dS=\frac{\partial S}{\partial T}dT + \frac{\partial S}{\partial E^2} d(E^2).
\end{equation}
Thus it is found that
\begin{equation}\label{tipoAB}
	dS=A(T,E^2)dT+B(T,E^2)d(E^2),
\end{equation}
where $A$ and $B$ are both functions of the two variables $T$ and $E^2$. The condition that $dS$ is a total differential requires that
\begin{equation}\label{totdiff}
	\frac{\partial B}{\partial T}=\frac{\partial^2 S}{\partial T \partial (E^2)}=\frac{\partial A}{\partial (E^2)}.
\end{equation}
Inserting $dQ$ from (\ref{dU_ext}) into relationships (\ref{leggi1}) and (\ref{leggi2}), we obtain
\begin{equation}\label{dS2}
	dS=\frac{1}{T}\left(\frac{\partial U}{\partial T}-\e_0 E^2 \frac{\partial\e_s}{\partial T}\right)dT+\frac{1}{T}\left(\frac{\partial U}{\partial (E^2)}-\frac{\e_0\e_s}{2} \right).
\end{equation}
This equation is of the type (\ref{tipoAB}); Eq. (\ref{totdiff}), therefore, becomes
\begin{displaymath}
	\frac{\partial}{\partial T}\left[\frac{1}{T}\left(\frac{\partial U}{\partial (E^2)}-\frac{\e_0\e_s}{2}\right)\right]=\frac{\partial}{\partial (E^2)}\left[\frac{1}{T}\left( \frac{\partial U}{\partial T}-\e_0 E^2 \frac{\partial \e_s}{\partial T}\right) \right],
\end{displaymath}
whose differentiation gives
\begin{displaymath}
	\frac{\partial U}{\partial (E^2)}=\frac{\e_0}{2}\left(\e_s+T\frac{\partial \e_s}{\partial T} \right).
\end{displaymath}
Integrating with respect to $E^2$ yields the energy density
\begin{equation}\label{energia}
	U=U_0(T)+\left(\e_s+T\frac{\partial\e_s}{\partial T} \right)\frac{\e_0 E^2}{2},
\end{equation}
where $U_0(T)$ is independent of $E^2$ but depends on $T$, thus it represents the energy of the dielectric in absence of a field.

From these relationships, Fr\"{o}hlich easily calculates entropy $S$: comparing Eq. (\ref{dS2}) with Eq. (\ref{dS1}), both $\partial S/\partial T$ and $\partial S/\partial (E^2)$ are known if $U$ is introduced from Eq. (\ref{energia}). Thus
\begin{gather}
	\frac{\partial S}{\partial T}=\frac{1}{T}\frac{\partial U_0}{\partial T}+\frac{\e_0 E^2}{2}\frac{\partial^2 \e_s}{\partial T^2},\nonumber\\
	\frac{\partial S}{\partial (E^2)}=\frac{\e_0}{2}\frac{\partial \e_s}{\partial T},\nonumber
\end{gather}
or integrating
\begin{equation}\label{SE1}
	S=S_0(T)+\frac{\e_0 E^2}{2}\frac{\partial \e_s}{\partial T},
\end{equation}
where $S_0(T)$ is the entropy in the absence of a field. From Eq. (\ref{FHelm}) one finally finds for the Helmholtz free energy
\begin{equation}\label{FHelm2}
	F=F_0(T)+\frac{\e_0\e_s E^2}{2},
\end{equation}
where $F_0(T)$ is the free energy in the absence of a field. We can summarize Fr\"{o}hlich's results by saying that under the above-stressed assumptions, relationships (\ref{RelGenerali}) are expressed by
\begin{subequations}
	\label{RelFr}
	\begin{align}
		U(T,E) & =U_0(T)+\frac{1}{2}\left(\e+T\frac{\de\e}{\de T}\right)E^2,\label{RelU_Fr}\\
		S(T,E) & =S_0(T)+\frac{1}{2}\frac{\de\e}{\de T} E^2,\label{RelS_Fr}\\
		F(T,E) & =F_0(T)+\frac{1}{2}\e E^2,\label{RelF_Fr}
	\end{align}
\end{subequations}
which are the \emph{Fr\"{o}hlich's thermodynamic relationships} \cite{Frohlich1958}, i.e. respectively \emph{Fr\"{o}hlich total energy}, \emph{Fr\"{o}hlich entropy}, \emph{Fr\"{o}hlich free energy} (all for volume unit) \cite{Parravicini2016,Parravicini2017,Tan2019}. We note that Relationships. (\ref{RelFr}) show that the standard expression (\ref{Densita2}) actually does provide the free energy only. Relations (\ref{RelFr}) have a very transparent and simple mathematical form, being the free energy proportional to the dielectric susceptibility and the entropy proportional to its derivative. However, the elegant formulation of (\ref{RelFr}) requires the above-mentioned specific hypotheses. We note that the requirement of considering an independent-on-time field (static condition) is a consequence of the reversibility assumption. On the other hand, the dielectric assumptions concern the specific features of the dielectric response, which is determined by the considered material. Actually, in the present discussion the key physical quantity is the entropy, which, we write by combining Eq. (\ref{SE1}) with Eq. (\ref{RelS_I}), i.e., in scalar approximation

\begin{displaymath}
	S(T,E)=S_0(T)+S_E(T,E)=S_0(T)+\frac{1}{2}\frac{\de\e}{\de T} E^2.
\end{displaymath}
Therefore, the field dependent part of entropy is
\begin{equation}\label{S_E}
	S_E(T,E)=\frac{1}{2}\frac{\de\e}{\de T} E^2.
\end{equation}
Relationship (\ref{S_E}) points out the \emph{Fr\"{o}hlich entropy} (FE), which can be seen as the entropy variation of the system due to the application of the electric field $E$, for volume unit \cite{Frohlich1958}. We will see that, for practical uses, from Eq. (\ref{S_E}) it is useful giving $S_E(T)$ for a unitary field \cite{Parravicini2016,Parravicini2017,Huang2019,Huang2020}, i.e.
\begin{equation}\label{s}
	s(T)=\frac{1}{2}\frac{\de\e}{\de T},
\end{equation}
which holds, of course, in the same hypotheses of Eqs. (\ref{RelFr}). $s(T)$ has the advantages of not depending on the strength of the electric field, featuring the structural information given by the dielectric response. Actually, Eq. (\ref{s}) is the \emph{key relationship} to exploit FE evaluation technique on the most of the physical systems.

\subsection{Generalized relationships: anisotropic and nonlinear dielectrics}
The constraints of linear and scalar dielectric response can be bypassed without specific assumptions on the physics of the considered material, but just through suitable mathematical treatments \cite{Nye1985,Landau}, which were recently developed by J. Parravicini \cite{Parravicini2018}. We here report the final obtained relationships, without recalling the details of the calculations.

\subsubsection{Anisotropic media}
If an anisotropic dielectric is assumed, the dielectric function must be expressed by a tensor $\epst$, so the relation between $\E$ and $\D$ has the general form $\D=\epst\cdot \E$, i.e.
\begin{gather}\label{lineareDE_an}
	D_i=\sum_j\e_{ij} E_j
\end{gather}
%
for $i,j=1,2,3$, where the $x,y,z$ components of the field are labeled as $E_x=E_1$, $E_y=E_2$, $E_z=E_3$, and similarly for $\D$. So the generalized form of Eqs. (\ref{RelFr}) for anisotropic dielectrics is
\begin{subequations}
	\label{Rel}
	\begin{align}
		U=U_0(T)+\frac{1}{2}\sum_{ij}E_j\left(\e_{ij}+T\frac{\de \e_{ij}}{\de T} \right)E_i,\label{RelU}\\
		S=S_0(T)+\frac{1}{2}\sum_{ij}\left(E_j \frac{\de \e_{ij}}{\de T} \right)E_i,\label{RelS}\\
		F=F_0(T)+ \frac{1}{2}\sum_{ij}\left(\e_{ij}E_j\right)E_i.\label{RelF}
	\end{align}
\end{subequations}
If we then focus on the entropy, from relationship (\ref{RelS}) we can evidently extend Eqs. (\ref{S_E}) and (\ref{s}) to the anisotropic case. So, Eq. (\ref{S_E}) becomes
\begin{equation}\label{S_E_anisotropa}
	S_E(T)=\frac{1}{2}\sum_{ij}\left(E_j \frac{\de \e_{ij}}{\de T} \right)E_i
\end{equation}
and Eq. (\ref{s}) will be
\begin{equation}\label{s_direzionale}
	s(T)\equiv \frac{1}{2}\sum_{ij}\left(\frac{\de \e_{ij}}{\de T} \right),
\end{equation}
which have, of course, the same physical meaning of their corresponding formulae in the scalar case. We note that the expression (\ref{S_E_anisotropa}) giving the field-induced entropy in an anisotropic medium is very similar to the usual expression giving the electric (free) energy density in the same conditions. This similarity can be highlighted by writing relationship (\ref{S_E_anisotropa}) in the form
\begin{equation}
	S_E(T)= \frac{1}{2}\frac{\de\D}{\de T}\cdot\E.
\end{equation}
This last one can be immediately compared with Eq. (\ref{Densita}), which we here write by underlining its dependence (through $\D$) on temperature, i.e. 
\begin{displaymath}
	U_e(T)= \frac{1}{2}\D\cdot\E.
\end{displaymath}

The physical meaning of relationships (\ref{Rel}) can be pointed out by providing formulae (\ref{Rel}) in the reference system of the principal axes \cite{Nye1985}, where tensor $\e_{ij}$ is diagonal, i.e.
\begin{equation}\label{TensoreDiagonale}
	\e_{ij}=
	\left(\begin{array}{ccc}
		\e_{11} &   0       &   0\\
		0       &   \e_{22} &   0\\
		0       &   0       &   \e_{33}
	\end{array}\right).
\end{equation}
Then, let's assume that the electric field is applied along the direction 1, where the system response is expressed by $\e_{11}$, i.e. $\E\equiv(E_1, 0, 0)$; so, the corresponding entropy variation will be
\begin{displaymath}
	S_E=S_{E_1}=\frac{1}{2}\frac{\de\e_{11}}{\de T},
\end{displaymath}
and analogously for a field along 2 and 3 directions. It is useful expressing Eqs. (\ref{Rel}) for the diagonalized system given by (\ref{TensoreDiagonale}) in polar coordinates. At a given fixed temperature $T$, we have
\begin{subequations}
	\label{Tuttipolari}
	\begin{align}
		U_E(E,\theta,\phi)=&\frac{E^2}{2}\Bigg[ \Big(\e_{11}+T\frac{\de \e_{11}}{\de T}\Big)\sin^2\theta\cos^2\phi +\nonumber \\
		+\Big(\e_{22}+T\frac{\de \e_{22}}{\de T}\Big)& \sin^2\theta\sin^2\phi +\Big(\e_{33}+T\frac{\de \e_{33}}{\de T}\Big)\cos^2\theta \Bigg],\\
		S_E(E,\theta,\phi)= &\frac{E^2}{2}\Bigg( \frac{\de \e_{11}}{\de T} \sin^2\theta\cos^2\phi +\nonumber\\
		+\frac{\de \e_{22}}{\de T} & \sin^2\theta\sin^2\phi + \frac{\de \e_{33}}{\de T}\cos^2\theta \Bigg),\\
		F_E(E,\theta,\phi)=&\frac{E^2}{2}\big( \e_{11} \sin^2\theta\cos^2\phi + \nonumber\\
		+  \e_{22}& \sin^2\theta\sin^2\phi + \e_{33}\cos^2\theta \big),
	\end{align}
\end{subequations}
where $E$ is the strength of the electric field. These relationships show that, if the electric field has strength $E$ and direction given by $\theta$ and $\phi$ angles, when $\e_{ij}$ has the diagonal form (\ref{TensoreDiagonale}), the components of the tensors $(\e_{ij}+T\de \e_{ij}/\de T)/2$, $(\de \e_{ij}/\de T)/2$, $\e_{ij}/2$ (at a given temperature) will be the coefficients for calculating the values of the variations of the considered thermodynamic quantities. On the other hand, they can also be seen as the resulting total energy, entropy, free energy, respectively, due to the application of a unitary field along one of the principal directions of $\e_{ij}$ \cite{Nye1985,Parravicini2016,Parravicini2017,Huang2019,Huang2020}. We note that, once the components of $(\e_{ij}+T\de \e_{ij}/\de T)/2$, $(\de \e_{ij}/\de T)/2$, $\e_{ij}/2$ are known along the principal directions (i.e. where $\e_{ij}$ is diagonal), relationships (\ref{Tuttipolari}) provide the values of the induced Fr\"{o}hlich's total energy, entropy, free energy for any direction of the considered electric field \cite{Nye1985,Devonshire1954}.

%
\subsubsection{Nonlinear media}
If the response of the dielectric to the field is not linear, second and third order nonlinear terms should be taken into account. So the field-dependent terms of (\ref{RelGenerali}) can be written in the form
\begin{subequations}
	\label{RelNL}
	\begin{align}
		U_E & =U_E^{(1)}+U_E^{(2)}+U_E^{(3)},\label{RelU_NL}\\
		S_E & =S_E^{(1)}+S_E^{(2)}+S_E^{(3)},\label{RelS_NL}\\
		F_E & =F_E^{(1)}+F_E^{(2)}+F_E^{(3)},\label{RelF_NL}
	\end{align}
\end{subequations}
where $U_E^{(1)}$, $U_E^{(2)}$, $U_E^{(3)}$ are, respectively, the term depending on linear, quadratic, and cubic susceptibilities, and analogously for entropy and free energy.

The fundamental relationship between the dielectric displacement $\D$ and the electric field $\E$ (Eq. (\ref{lineareDE_an})) are
\begin{equation}\label{D2_anisotropa}
	D_i=\sum_j \e^{(1)}_{ij} E_j + \sum_{jk} \e^{(2)}_{ijk}E_j E_k + \sum_{jkl}\e^{(3)}_{ijkl}E_j E_k E_l,
\end{equation}
where $\e_{ij}^{(1)}=\e_{ij}$ of (\ref{lineareDE_an}),  $\e_{ijk}^{(2)}$ and $\e^{(3)}_{ijkl}$, for $i,j,k,l=1,2,3$, are respectively the absolute second- and third- order nonlinear dielectric functions.  The expression of the total energy, in the form of (\ref{RelU_I}), for a nonlinear medium is

\begin{gather}
	U=U_0(T)+\frac{1}{2} \sum_{ij}\left(\e^{(1)}_{ij}+T\frac{\de\e^{(1)}_{ij}}{\de T}\right)E_i E_j+\nonumber\\
	+\frac{1}{3}\sum_{ijk}\left(\e^{(2)}_{ijk}+T\frac{\de\e^{(2)}_{ijk}}{\de T}\right)E_i E_j E_k +\nonumber\\
	+\frac{1}{4}\sum_{ijkl}\left(\e^{(3)}_{ijkl}+T\frac{\de \e^{(3)}_{ijkl}}{\de T}\right)E_i E_j E_k E_l\label{RelU2tens},
\end{gather}
while the entropy
\begin{gather}
	S=S_0(T)+\frac{1}{2} \sum_{ij}\left(\frac{\de\e^{(1)}_{ij}}{\de T}\right)E_i E_j+\nonumber\\
	+\frac{1}{3}\sum_{ijk}\left(\frac{\de\e^{(2)}_{ijk}}{\de T}\right)E_i E_j E_k +\nonumber\\
	+\frac{1}{4}\sum_{ijkl}\left(\frac{\de\e^{(3)}_{ijkl}}{\de T}\right)E_i E_j E_k E_l,\label{RelS2tens}
\end{gather}
and the Helmholtz free energy:
\begin{gather}
	F=F_0(T)+\frac{1}{2} \sum_{ij}\left(\e^{(1)}_{ij}\right)E_i E_j+\nonumber\\
	+\frac{1}{3}\sum_{ijk}\left(\e^{(2)}_{ijk}\right)E_i E_j E_k +\nonumber\\
	+\frac{1}{4}\sum_{ijkl}\left(\e^{(3)}_{ijkl}\right)E_i E_j E_k E_l. \label{RelF2tens}
\end{gather}
\noindent These last expressions have the form of (\ref{RelGenerali}) and their field-dependent parts have the form of (\ref{RelNL}). They are the most general formulation of the Fr\"{o}hlich relationships for a material with both second and third order nonlinearities. The scalar formulation of (\ref{RelU2tens}), (\ref{RelS2tens}), and (\ref{RelF2tens}) can be useful for highlighting their physical meaning. It is
\begin{subequations}
	\label{Rel2isotr}
	\begin{align}
		U=U_0(T)+\frac{1}{2}\left(\e^{(1)}+T\frac{\de\e^{(1)}}{\de T}\right)E^2+\nonumber\\
		+\frac{1}{3}\left(\e^{(2)}+T\frac{\de\e^{(2)}}{\de T} \right)E^3 +\frac{1}{4}\left(\e^{(3)}+T\frac{\de\e^{(3)}}{\de T} \right)E^4,\label{RelU2}\\
		S=S_0(T)+\frac{1}{2}\frac{\de\e^{(1)}}{\de T} E^2 + \frac{1}{3}\frac{\de\e^{(2)}}{\de T}E^3+\frac{1}{4}\frac{\de\e^{(3)}}{\de T}E^4,\label{RelS2}\\
		F=F_0(T)+\frac{1}{2}\e^{(1)}E^2 + \frac{1}{3}\e^{(2)}E^3 + \frac{1}{4}\e^{(3)}E^4\label{RelF2}.
	\end{align}
\end{subequations}
\noindent These equations point out that a not-null second-order nonlinearity makes the thermodynamic quantities depending on the cube of the applied field: this implies that a dielectric system with $\e^{(2)}\neq 0$ must be noncentrosymmetric, showing that a material with a not-null static second-order nonlinearity necessarily hosts a breaking of symmetry.

\section{Physical meaning of Fr\"{o}hlich entropy}
We here examine three different interpretations of Fr\"{o}hlich entropy. The \emph{first} one is correlated with the simplest model of a dielectric, i.e. a set of dipoles that may be oriented by a measuring field. The \emph{second} one stresses that FE can be used to check the state of order of a system; namely, it is found very useful to describe the features of phases and correlated transition processes. The \emph{third} framework highlights the capability to give a measure of the stability within an explored phase. All these views have been practically exploited in the investigation of physical systems where FE evaluation has been carried out. It is worth noting that the original previously quoted Fr\"{o}hlich's interpretation\footnote{Section \ref{INTRO}, pg. \pageref{citazione1}} was also recalled e.g. by Scaife and by B\"{o}ttcher, with similar sentences, in their classical monograph on dielectrics \cite{Scaife1989,Scaife1998,Boettcher1973}\footnote{See, in particular, chapter 3 \S13 of \cite{Boettcher1973}.}.

\subsection{Fr\"{o}hlich entropy as a figure of merit of orientational disorder}\label{PhysMean:Interp}
\noindent The relationships involving $U$, $S$, and $F$ exposed in the previous sections have deep physical meanings. In fact, Eq. (\ref{FHelm2}) for the free energy shows that the amount of electric energy available in an isothermal reversible process is $\e_0\e_s E^2 /2$. Through Eq. (\ref{energia}), on the other hand, Fr\"{o}hlich demonstrates that substances following the Curie-Weiss law, such as e.g. diluted dipolar gases, the temperature dependent part of $\e_s$ does not provide any contribution to the total energy \cite{Frohlich1958}\footnote{Pg. 12, \S3, Eqs. (3.11) and (3.14)}.

\noindent For our treatise, the most important results involve the entropy as given by Eq. (\ref{S_E}). It shows that the entropy of the considered system is increased by the field if $\partial\e_s/\partial T$ is positive, while it is decreased when this quantity is negative. We follow the original Fr\"{o}hlich's rationale, which was exposed in the above-quoted sentences from \cite{Frohlich1958}. Let us consider a physical system with dipoles being free-to-reorient, such as liquids and gases. In these conditions $\e_s$ decreases with increasing $T$ \cite{Johari2013}: here, an external field forces the dipoles to coherently reorient, so creating ordering. This may be expected because the field will orientate some of the dipoles which in the absence of a field are at random. Conversely, if we consider a system where inherent dipoles are in a well ordered state -such as typically found in a crystalline solid- here they lie in stable energy minima. The correlated trend of $\e_s$ is to increase with $T$. This means that an external field increases the disorder. This is understandable if one assumes that in the absence of a field the dipoles are in a well-ordered state as may be expected in many solids. The effect of the field is to perturb a (relatively) stable configuration by turning some of the dipoles into different directions, so the field can only decreases the existing order. Therefore FE can be considered as a a measure of the orientation degree, i.e. an \emph{orientational entropy}, providing information about the rearrangement of the oriented electric dipoles which belong to the system \cite{Sethna2006,Johari2016,Swiergel2011Orient}. An ideal behavior of $\e(T)$ and the correlated $S_E(T)$ are depicted in Fig. \ref{SchemaFrohlich}, which reproduces the scheme in the original Fr\"{o}hlich's treatise. Of course, this picture is not limited to specimens explicitly hosting dipoles (e.g. polar liquids \cite{Johari2013,Johari2016}), but it can be applied to all samples where the internal charge distribution can be suitable approximated to dipolar terms, at least at the first order. This means that this relatively simple physical interpretation holds for the greatest part of the condensed matter: actually, in his treatise Fr\"{o}hlich provides this interpretation in a very general and concise way \cite{Frohlich1958}. However, we recall that the above-mentioned hypotheses about the thermodynamic equilibrium, static conditions (no time dependent fields \cite{Johari2013,Richert2017}) and the suitable constitutive relationships between $E$ and $D$ (or $P$) are key factors. The strength of $E$, on the other hand, should be carefully handled, because when very high fields are applied (typically $\simeq 10^3$-$10^4$ V/cm), the system can be strongly perturbed and $E$ can affect not only the orientation of the internal dipoles, but their structure also (coercive field effects) \cite{Johari2013,Jona1993,Kao2004}. Differently, the probing electric field employed in the here discussed experiments is typically $\lesssim 5$ V/cm.%
%

\subsection{Fr\"{o}hlich entropy: a convenient state of order figure of merit}
\noindent We want here to emphasize that Fr\"{o}hlich entropy can be fruitfully exploited to finely describe the state of order (or disorder) of a macroscopic system. Moreover, since FE derives from dielectric measurements, it allows to almost continuously follow the evolution of the order/disorder state, which signals the features of the considered phase and correlated transitions. Namely, the powerful of FE evaluation is particularly evident when phase transitions occur. Actually, an order/order transition manifests itself as a variation of the positive slope of the $S_E$ curve, since the entropy contribution supplied by the field is positive both before and after the transition. The corresponding values of $S_E(T)$ are in accordance to the degree of order of each phase \cite{Frohlich1958,Scaife1998,Parravicini2003APL,Sassella2011,Parravicini2012}. In the transition interval, a characteristic peak can be found, whose width can be assumed to estimate the temperature extension of the transition process. An analogous behavior, with negative $S_E(T)$ values, is displayed in the case of a disorder/disorder transition. On the other hand, when an order/disorder transition occurs, a two-opposed-peaks lineshape of $\partial \e_s/\partial T$ is typically displayed (Fig. \ref{SchemaFrohlich}), in correspondence with the positive and negative slope in the related permittivity peak, with negative part marking a temporary disordered state in the material. Therefore, the estimation of FE is demonstrated to be a particularly sensitive probe for investigating phase transition phenomena. These features can be found in a very large class of systems.

\subsection{Fr\"{o}hlich entropy: a quantitative evaluation of phase stability}
FE was also demonstrated to be able to provide specific information about the stability of a considered long-range phase. This conceptual approach was developed mainly in works concerning organic molecular crystals \cite{Sassella2011,Parravicini2012}. This concept starts from the above recalled Fr\"{o}hlich's interpretation (Sec. \ref{PhysMean:Interp}). When an electric field is applied, if the corresponding value of $|S_E(T)|$ is great, this implies that the system is strongly modified by the application of the $E$. On the other hand, if the corresponding value of $|S_E(T)|$ is relatively small, this means that $E$ slightly changes the status of the system. Then, we ca reasonably assume that, in the considered conditions, the system in the first case is in a less stable configuration than in the second case. So, both for ordered ($S_E<0$) and disordered ($S_E>0$) configurations, FE is able to provide an immediate quantitative estimation of the stability of a given temperature dependent phase and to compare this feature between different phases. In particular, a nearly constant trend of $S_E(T)$ points out a temperature region with highly stable features, while a steeply changing of $S_E(T)$ is be associated to instability features, which typically become the most marked in correspondence to the phase transition temperature regions.

\noindent The scheme in Fig. \ref{SchemaFrohlich} can be useful to grasp this physical picture. We can see that the greatest values of $|S_E(T)|$ correspond to the region around the phase transition, where the system configuration is reconfiguring and its long-range state is not stable. As more as it moves away from the transition, the phase is stabilizing, so the applied electric field only negligibly perturbs the system: this is pointed out by the small values of $|S_E(T)|$, which are decreasing slower and slower where a long-range phase is achieved. Moreover, a further signature of phase instability is given by the arising of oscillations of $S_E(T)$ as a function of temperature (see e.g. Fig. \ref{FigParravicini2012-01}b below), where a regular (nearly constant) trend of FE (see e.g. Fig. \ref{FigSassella2011}b below) is instead a clear signal of a highly stable and not changing phase \cite{Sassella2011,Parravicini2012}. Further signatures of instability phenomena are the presence of oscillations of $S_E(T)$ \cite{Sassella2011}, showing that the dipoles of the considered system are quickly reconfiguring. In particular, when these oscillations cross the $S_E=0$ value, the presence of oscillations signals the arising of hybrid and metastable phases, where order and disorder coexist. There, the order degree of the system is reconfiguring, rapidly passing from disordered to ordered state, or it is not well defined (Fig. \ref{FigDelRe2015NP}).
\begin{figure}
	\begin{center}
		\caption{From Supplementary Information of \cite{DelRe2015} - (a) Fr\"{o}hlich entropy ($S_E$ is here indicated with $\Delta S$, per volume unit, normalized to the squared probe-field amplitude $E^2$) in proximity of the Curie point of a disordered potassium-based perovskite crystal and (b) blow-up of shaded region illustrating phase fluctuations from liquid-like to solid in the heating stage.}
		\label{FigDelRe2015NP}
	\end{center}
\end{figure}
%
%
%
%


\section{Investigation of physical systems by using Fr\"{o}hlich entropy}\label{Investigation}
\subsection{General view}
The generality of the relationship that defines Fr\"{o}hlich entropy and the use of his interpretation in studying very different physical systems raise the problem of practically satisfying its theoretical validity conditions, whose assumptions are listed in section \ref{SezioneBasic} (pg. \pageref{F_assump}). These guarantee that the formalism of relation (\ref{RelS_Fr}) (and its extensions (\ref{RelS}) and (\ref{RelS2tens})) holds. Experiments demonstrated that Fr\"{o}hlich entropy relationship can be applied in less stringent conditions with respect to the formal ones. Namely, relationship (\ref{RelS_Fr}) was experimentally found to hold in quasi-equilibrium states \cite{Parravicini2016,Sassella2011,Parravicini2012} and quasi-static conditions \cite{Parravicini2003APL,Parravicini2003EPJD,Parravicini2003PSSB,Samanta2016}, i.e. they still hold when relaxation phenomena are negligible: this implies that quasi-static fields (frequencies until $\simeq$ 100 MHz \cite{Jadzyn2007}) can be handled. On the other hand, the validity of hypotheses on dielectric features strongly depends on the investigated material. Namely, the isotropy assumption of Fr\"{o}hlich discussion can be not assumed, so the investigation of systems whose response is intrinsically directional is also possible. In fact, it has been accomplished on anisotropic crystals \cite{Parravicini2016,Parravicini2017,Huang2020} and nematic compounds \cite{Jadzyn2007,Jadzyn2008_01,Jadzyn2008_03,Jadzyn2009,Jadzyn2007_02}, also before the mathematical formulation of the anisotropic extension of Fr\"{o}hlich's original relationship \cite{Parravicini2018}, which are here reported in Eqs. (\ref{RelS}) and (\ref{RelS2tens}). Although the extension of Eq. (\ref{RelS_Fr}) to nonlinear dielectric response has been already provided (i.e. high electric field conditions), no experimental works explicitly involving nonlinear regimes has been up today already carried out.

We here provide a comprehensive landscape of the physical systems where FE has been employed as an investigation instrument aimed at gaining specific information about material state and its evolution. We will not discuss investigations where, although Fr\"{o}hlich relationships are used, experimental conditions determine that the application of the electric field not negligibly modify the structure of the considered physical system. In other words, we will consider those works where the applied $E$ can be considered a \emph{probing field}. For this reason, we will not discuss the works of the groups of G. Johari and R. Richert about structural polar liquids, where relationships involving FE are employed for describing structural changes induced by the application of the electric field \cite{Johari2013,Johari2013_02,Johari2016,Samanta2015,Samanta2016,Kaminska2016}.

First we will consider the investigation of phase transitions and melting processes in confined metallic Ga nanoparticles. They are the first system where FE evaluation has been systematically exploited for interpreting experimental results by GB. Parravicini et al. of Pavia group in 2003 \cite{Parravicini2003APL,Parravicini2003EPJD,Parravicini2003PSSB}. Then, we will discuss systems of polar liquids and glasses, where FE has been employed in several works since 2007, mainly due to J. Jadz\.{y}n et al. of Pozna\'{n} group. Subsequently, the use of FE for analyzing phase transitions and phase stability in semiconductor molecular crystals is considered. Fr\"{o}hlich's argument has been fruitfully used in the investigation of disordered ferroelectric crystals also, whose evolution of the order state has been finely described and a directional description has been remarkably provided. Finally, works on FE in polymeric films and diluted enzymes will be discussed.

\subsection{Confined metallic nanoparticles}\label{Ga_nanopart}
As hinted before, the first class of physical systems where the Fr\"{o}hlich formalism was applied in the interpretation of experimental results are confined metallic nanoparticles. Specifically, Ga nanoparticles embedded in a SiO$_x$ matrix are considered, a paradigmatic arrangement analogous to other confined systems where the interfaces play a key role \cite{Nisoli1997,Konrad1998}.

The issue of phase transitions in nanometric-size metallic particles studied through dielectric spectroscopy had been faced in a preliminary work by Pavia group in 2000 \cite{Tognini2000}. The investigation of Ga was considered a very interesting topic due to the rich polymorphism of this element; moreover, several investigations gave the evidence of the existence, in submicrometer particles, of different metastable structural phases (named $\beta$, $\gamma$, $\delta$, and $\epsilon$) \cite{Bernasconi1995,Konrad1998,Nisoli1997,Bosio1973, Parravicini2006}. The main result of this experimental work was the assessment of hysteresis loop crossing melting and solidification temperatures. This investigation, moreover, demonstrated that the employed dielectric technique was able to inspect the various solid phases of the confined Ga nanoparticles \cite{Tognini2000}. 

After this work, a specific effort was accomplished towards getting more information about the nanoparticle phase evolution from dielectric measurements. Hence, a new analysis approach exploiting Fr\"{o}hlich's theory and interpretation was given in 2003, in a work where melting and premelting processes are investigated ``through capacitance measurements'' \cite{Parravicini2003APL}. The Fr\"{o}hlich's interpretation was more specifically applied in the following two works, where the temperature evolution of the ``degree of order'' \cite{Parravicini2003PSSB} and the temperature-dependent order-disorder changes \cite{Parravicini2003EPJD} are analyzed. In these works the capacitance measurements on metallic nanosystems are proved to give detailed information on the phase transitions and the correlated premelting and melting processes, with their dependence on sizes of involved particles \cite{Parravicini2003APL,Parravicini2003EPJD,Parravicini2003PSSB,Parravicini2004}. 

The investigated samples containing Ga particle layers may be considered as a system of two capacitors in series, the layer of SiO$_x$ and the layer of Ga clusters, following Fig. \ref{FigParravicini2004APL} from \cite{Parravicini2004APL}. In this reported picture the typical experimental arrangement is depicted: being the employed configuration a plane-parallel capacitor, the derivative of the capacitance $C(T)$ is proportional to FE given by Eq. (\ref{RelS_Fr}), i.e. $dC(T)/dT\propto S_E(T)$.

\begin{figure}
	\begin{center}
		\caption{From \cite{Parravicini2004APL} - Typical sample arrangement of the works on Ga confined nanoparticles here considered. (a) AFM plain-view image of the Ga nanoparticles arrays in sample; (b) schematic representation of the sequence SiO$_x$/Ga nanoparticles layer/SiO$_x$.}
		\label{FigParravicini2004APL}
	\end{center}
\end{figure}

\subsubsection{Phase transitions}
The features of phase transitions are investigated on two different samples of Ga nanoparticles, whose the most representative are particles of 20 nm radius (Ga20). Preliminary transmission electron microscope (TEM) measurements as a function of temperature, which are reported in Fig. \ref{FigParravicini2003PSSB_01}, indicate signature of an evident transition \cite{Parravicini2003PSSB}. Actually, the data of TEM on Ga20 sample point out a solid-liquid transition of the Ga particles at the temperature $T_c = 250$ K. Specifically, the crystalline phase in solid state is identified by TEM as $\delta$-phase displaying trigonal structure, in very good agreement with the data reported in literature. Remarkably, when the detected transition from solid to liquid phase is considered, the melting process is found to start about 65 K below the full melting temperature value.
\begin{figure}
	\begin{center}
		\caption{From \cite{Parravicini2003PSSB} - TEM measurements - (a) Ga $\delta$ phase: selected area of electronic diffraction (SAED) radial plot of Ga clusters in the solid phase. In the upper part of the figure the calculated diffraction pattern of the Ga $\delta$ phase is reported. (b) The correspondent SAED pattern. The SAED patterns at c) 240 K and d) 250 K, respectively, giving evidence of solid-liquid phases and full liquid phase are reported.}
		\label{FigParravicini2003PSSB_01}
	\end{center}
\end{figure}
The $\e=\e(T)$ evolution is plotted in Fig. \ref{FigParravicini2003EPJD_A}: both real and imaginary part of the measured dielectric function show the occurrence of a transition centered at 185 K. It is worth stressing the high sensitivity of such a kind of measurements: actually the corresponding electronic diffraction measurements (Fig. \ref{FigParravicini2003PSSB_01}) are able to detect hints of a transition only at 240 K and its end at 250 K.

The FE correlated to the dielectric response of Fig. \ref{FigParravicini2003EPJD_A} is reported in Fig. \ref{FigParravicini2003EPJD_B}, where it is given in terms of derivative of the dielectric capacitance $C=C(T)$ as a function of the temperature: $dC/dT\propto S_E\propto\partial\e/\partial T$. Plot of Fig. \ref{FigParravicini2003EPJD_B} highlights that the change from a positive to negative value of $dC/dT$ corresponds to a transition from an ordered phase to a disordered one, i.e. from the solid to a liquid state of Ga. In particular, the shown trend of FE vs. temperature demonstrates that the disorder of the system increases with $T$ until 250 K. We note that the presence of a slope change at 220 K signals the arising of pretransitional features, while the occurring of the specific transition is abruptly pointed out by the change of sign at 185 K. Moreover, when the temperature is increased, the trend of $dC/dT$ (i.e. the FE) indicates the sequence of transitions through several metastable phases, in agreement with those reported in the literature \cite{Bernasconi1995,Konrad1998,Nisoli1997,Bosio1973}. As an apt conclusion of these studies, a comprehensive EXAFS investigation on Ga nanoparticles accurately confirmed the polymorphism and matastabilities highlighted by dielectric measurements and FE analysis \cite{Ghigna2007}.
\begin{figure}
	\begin{center}
		\caption{From \cite{Parravicini2003EPJD} - 20 nm size sample: capacitance and conductance of the Ga nanoparticles layer as a function of temperature in the range below the full melting temperature.}
		\label{FigParravicini2003EPJD_A}
	\end{center}
\end{figure}
\begin{figure}
	\begin{center}
		\caption{From \cite{Parravicini2003EPJD} - The derivative of the Ga layer capacitance $C$ of Fig. \ref{FigParravicini2003EPJD_A} as a function of the temperature $T$, versus temperature (solid line). The derivative is proportional to the entropy due to the applied field. The change of sign, at $T = 185$ K, evidences an order-disorder transition. The dotted line shows a linear behavior in the 185-220 K interval.}
		\label{FigParravicini2003EPJD_B}
	\end{center}
\end{figure}
%


\subsubsection{Melting processes}\label{MeltingGa}
We here discuss more deeply the specific item of melting process. As we hinted before, in low-size systems the amount of surface atoms with respect to the bulk ones can not be considered negligible, so the melting processes in such systems display peculiar features. In these conditions the use of FE analysis, which gives the possibility of continuously following the evolution vs. temperature, enables to carry out a detailed study. This topic was specifically faced in two works about the melting evolution in Ga nanoparticles \cite{Parravicini2003APL,Parravicini2003PSSB}.

Among the order/disorder phase transitions, melting is one of the most extensively observed, but achieving a complete microscopic description often is difficult. For instance, it is well known that in the considered Ga metallic nanoparticles thermodynamic properties related to melting show significant deviations with respect to the bulk ones. The key role of the surface in the melting processes was stressed in different experimental investigations \cite{Jin2001,Cahn2001}. By means of optical, Dark Field Electron Microscopy (DFEM) and electronic microscopy (TEM) techniques on metal nanoparticles evidence was given of premelting processes, namely the formation of a liquid layer, some degrees below the complete melting, that progressively expands with increasing $T$ (Fig. \ref{FigParravicini2003PSSB_01}). Melting precursor effects specifically involving the first surface atom layers were investigated (surface pre-roughening, roughening, reconstruction) \cite{Frenken1986}. This is particularly relevant in the nano-size scale, where the role of surfaces and interfaces becomes predominant \cite{Stella1996}. Carried out measurements highlight the occurrence of surface changes as a thin molten layer around the solid core which arises some degrees before full melting. This is due to the interplay of the involved interface quantities $\gamma_{sm}$, $\gamma_{sl}$, $\gamma_{lm}$, i.e. the free energies per unit area of solid-matrix (sm), liquid-matrix (lm) and solid-liquid (sl) interfaces respectively. The melting happens when the inequality $\gamma_{sm}-\gamma_{sl}-\gamma_{lm}>0$ holds. In Ga nanoparticles, when TEM investigation is accomplished, above-shown SAED pictures and spectra point out a complete bulk melting of the embedded nanoparticles between 240 and 250 K (Fig. \ref{FigParravicini2003PSSB_01}). Conversely, FE analyses are found to be able to specifically detect surface precursor phenomena. %
%
Actually, the dielectric response is closely related to the polarized surface of the nanoparticles: this implies that such a technique selectively gets the signal of first atomic layers of the considered physical system. %
%
%
%
\begin{figure}
	\begin{center}
		\caption{From \cite{Jin2001} - The experimental Lindemann parameter ($\delta_L$), the effective shear modulus $\Delta C_S$ (squares), and $\alpha_2$ parameter versus temperature (red circle added).}
		\label{FigJin2001_A}
	\end{center}
\end{figure}
\begin{figure}
	\begin{center}
		\caption{From \cite{Jin2001} - Calculated percentage of Lindemann particles as a function of temperature as also reported by \cite{Cahn2001}.}
		\label{FigJin2001_B}
	\end{center}
\end{figure}
%

Fig. \ref{FigParravicini2003EPJD_B} displays the derivative of $C(T)$, which is $dC(T)/dT\propto S_E(T)$. The plot clearly points out the sign change at 185 K, assessing the onset of an order-disorder transition. In the interval 220-250 K we compare TEM images with FE analysis results. These are characterized by a progressively more pronounced slope, as one expects, when the instabilities are strongly enhanced and start involving the inner part of the system \cite{Frohlich1958,Parravicini2003APL,Parravicini2003PSSB}. In this temperature interval, when inner melting is occurring, the slope manifests a clear divergence. In Fig. \ref{FigParravicini2003EPJD_B} a linear slope is observed in the $T$-range 185-220 K, where TEM data are not modified by any $T$-dependent process. The data can be interpreted in terms of onset of vibrational instabilities at the very surface before reaching the critical value. This value is predicted for melting by the Lindemann criterion, expressed by the Lindemann ratio, $\delta_L$, i.e. the ratio between the \emph{root-mean-square displacement}, rmsd, of a particle from its equilibrium position to the distance between them. It is worth noting  that the presence of two regimes had been predicted in a similar context by molecular dynamic (MD) simulations performed on a nanosystem made of 6912 (a size comparable to metallic nanoparticles here investigated) obeying Lennard-Jones potential and overheated using appropriated step-by-step procedures \cite{Jin2001,Cahn2001}. In this vein, it is illuminating to compare Fig. \ref{FigParravicini2003EPJD_B} with Figs. \ref{FigJin2001_A} and \ref{FigJin2001_B} (from \cite{Jin2001}): the entropy divergence compares nicely to the divergence of Lindemann atoms vs. temperature. The former appears to be smoother, since it takes place in real crystals, with evidence of surface melting. The linear slope of the FE variation from 185 K to about 220 K can be related to the linear increase of the rmsd of (surface) atoms before reaching the inner melting region (see Figs. \ref{FigJin2001_A} and \ref{FigJin2001_B}), in agreement with \cite{Jin2001} . We stress that the very high specific sensitivity to the local surface atoms evolution of the dielectric measurements allows us to monitor the different stages of pre-melting and full melting processes in a wide temperature region. The existence of two regimes pointed out in the $dC/dT$ trend is an important result illustrating the potential of the employed method.

%
%

\subsubsection{Summary}
Summarizing, the investigation of phase transitions on Ga nanoparticles shows that the FE evaluation method is particularly sensitive to the evolution of the low-sized (less than $\mu$m) systems, where the surface features are a key ingredient to produce melting and, more generally, phase transition processes. The relationships given by Fr\"{o}hlich between dielectric function and entropy is demonstrated to be useful in investigating surface phenomena in these nanometric systems in terms of order-disorder transitions and their evolution. The FE estimation, although it is an intrinsically macroscopic measurement, is found to be very accurate in describing the dynamics of surface atoms, allowing to follow the different phases of premelting and full melting. Therefore, the investigation of melting processes by dielectric measurements and related FE estimation was demonstrated to display the strict correlation of the dielectric response with the surface polarization, notably the surface first atomic layers and a high sensitivity for finely following the phase transition evolution.

\subsection{Polar liquids and glasses}
Further wide classes of materials where FE interpretation was demonstrated to be a useful experimental tool is that of polar liquids and glasses. Frohlich's relationships were used to describe both structural liquids and nemoticons, in a series of works mainly carried out by J. Jad\.{z}yn and his group.

\subsubsection{Structural liquids and glasses}
FE evaluation is exploited to describe the arrangement of structural liquids, both organic and inorganic, in suitable solvents. Studies are carried out on polar inorganic pure formamide %
\cite{Jadzyn2012}, propylene carbonate and dimethylsulfoxide (DMSO) with their mixtures \cite{Plowas2016}, highly-polar amide compounds such as n-methylpropionamidine%
, its homologue n-methylacetamide, and n-ethylacetamide %
%
\cite{Swiergel2011Orient}, organic lactones %
\cite{Swiergel2012}, inorganic lithium perclorate in polyethylene glycols solvent \cite{Swiergel2015}. On a different research line, a theoretical study of D.V. Matyushov et al. takes into account the configurational entropy of several polar organic liquids undergoing glass transition \cite{Matyushov2016}.

A meaningful example of such a kind of studies is reported in Fig. \ref{FigSwiergel2011Fig9-Fig10}, where the FE behavior of n-methylpropionamidine (NMP) and n-ethylacetamide (NEA) is depicted \cite{Swiergel2011Orient}. It is calculated by the relationship:
\begin{equation}\label{Jadzyn1}
	\frac{\Delta S}{E^2}=\frac{S(E,T)-S_0(T)}{E^2},
\end{equation}
where $\Delta S$ is the entropy variation for unit of volume which is due to the application of the electric field\footnote{$\Delta S$ of Eq. (\ref{Jadzyn1}) is the same quantity of $S_E$ in Eq. (\ref{S_E})}, so $\Delta S/E^2$ is $s$ of Eq. (\ref{s}). Here FE evaluation is exploited to obtain information on the intermolecular interactions and the molecular self-aggregation abilities of the considered liquids.
\begin{figure}
	\begin{center}
		\caption{From \cite{Swiergel2011Orient} - (a) temperature dependence of the static dielectric permittivity of n-methylpropionamidine (NMP), n-ethylacetamide (NEA) experimentally obtained in \cite{Swiergel2011Orient} (full points) and the literature data. Permittivity data for n-methylacetamide NMA are also depicted for comparison. (b) correlated temperature dependence of the orientational entropy variation for n-methylpropionamidine (NMP), n-ethylacetamide (NEA) and n-methylacetamide NMA. Data for dimethylsulfoxide (DMSO), which is not self-associated, are also reported for comparison.}
		\label{FigSwiergel2011Fig9-Fig10}
	\end{center}
\end{figure}
%
%
The plot in Fig. \ref{FigSwiergel2011Fig9-Fig10}b shows, for all the studied compounds, that the FE is negative, as we expect for a disordered dipolar liquid, where an applied field creates order. When the temperature is increased, the lowering of the modulus of the induced negative entropy gives a measure of the progressive decrease of the disorder in the samples. Actually, the authors associate this response to the formation of multimolecular aggregations which continuously increase their order. The comparison with the trend of non-self-associated compounds such as DMSO, whose FE is constant vs. temperature, confirms that FE response describes these peculiar behavior of the dipolar molecules \cite{Swiergel2011Orient}.

Such experimental approach was also used to specifically highlight the variation of orientational features as a function of solvent concentration: this was carried out, as hinted above, with mixtures of propylene carbonate and DMSO, a compound of significant application interest \cite{Plowas2016}. The plot of Fig. \ref{FigPlowas2016Fig5} shows FE dependence on both temperature and solvent concentration $x_{\text{DMSO}}$. It allows to follow the structural evolution of the physical system with a crossed-view between stoichiometry and temperature.
\begin{figure}
	\begin{center}
		\caption{From \cite{Plowas2016} - FE for mixtures of propylene carbonate and DMSO of different compositions at different temperatures.}
		\label{FigPlowas2016Fig5}
	\end{center}
\end{figure}

\subsubsection{Nematic solutions}
A further set of compounds where FE evaluation is fruitfully employed is liquid crystals. In particular, this approach was used in 2008 to describe the dynamics of mesogenic 4-(trans-4'-n-hexylcyclohexyl)isothiocyanatobenzene molecules in prenematic conditions \cite{Jadzyn2008_02}. Notably, the authors claimed that this work, and other of the same group \cite{Jadzyn2007,Jadzyn2007_02,Jadzyn2007_03}, were the first to experimentally exploit Fr\"{o}hlich equations and interpretation, although the previously discussed works on Ga nanoparticles had been published four years before (\cite{Parravicini2003APL,Parravicini2003EPJD,Parravicini2003PSSB}, section \ref{Ga_nanopart}). All these studies face organic compounds such as strongly polar alkylcyanobiphenyls and mesogenic liquids with different polarities, namely n-heptylcyanobiphenyl and 4-(trans-4'-n-hexylcyclohexyl)isothiocyanatobenzene, whose molecular and prenematic dynamics are investigated \cite{Jadzyn2007_02,Jadzyn2007_03}. Prenematic processes are also studied, through FE evaluation, in organic nonpolar compounds such as admixture of 4-n-ethylcyclohexyl-4'-n-nonylphenyl (C$_2$H$_5$C$_y$H$_x$P$_h$C$_9$H$_{19}$, 2C$_y$Ph9)
in mesogenic solvent of n-hexylcyanobiphenyl %
\cite{Jadzyn2008_02},4-(trans-4'-n-hexylcyclohexyl)isothiocyanatobenzene and 4-cyanophenyl-4'-n-heptylbenzoate \cite{Jadzyn2008_03,Jadzyn2009}, while in 4-cyanophenyl 4'-alkylbenzoates and 4-n-octyl(4'-cyanophenyl)benzoate the dipolar orientational and isotropy features are correlated to the molecular structures and dielectric properties \cite{Jadzyn2008_01,Bauman2010}.

In Fig. \ref{FigJadzynFig6_2008_02} the trend of FE as a function of temperature is displayed for different concentrations of a nonpolar admixture of 4-n-ethylcyclohexyl-4'-n-nonylphenyl in a mesogenic solvent \cite{Jadzyn2008_02}. This is calculated from the dielectric permittivity values reported in Fig. \ref{FigJadzynFig4_2008_02} by exploiting the relationship (\ref{Jadzyn1}) (identical to Eq. (\ref{s}), which is also recalled in \cite{Jadzyn2008_02}). The trends of dielectric function (Fig. \ref{FigJadzynFig4_2008_02}) and correlated FE (Fig. \ref{FigJadzynFig6_2008_02}) are assumed to describe the features of dipole orientation phase transitions, namely the isotropic-nematic (I-N) transition. Hence the critical I-N temperatures are indicated by a sharp increase both of $\e_s(T)$ and the $d\e_s/dT$ (dashed line). In other words, the FE plots accurately describe the pretransitional (prenematic) processes. At higher temperatures, $d\e_s/dT$ is negative: following Fr\"{o}hlich's interpretation, this shows a typical disordered polar liquid-like behavior. Lowering temperature a continuous decrease of the disorder is pointed out, also with the change of FE sign, so displaying a further increase of correlation among the molecules. This gradual increasing until the vertical trend of the plots indicates the full I-N transition. %
It is worth noting that the pure 2C$_y$Ph9 without solvent (which is indicated as $x=1$ and 2C$_y$Ph9 in the plots) shows no FE variation. We here note the specific sensitivity of FE evaluation to pretransitional processes, which is able to describe prenematic behavior in these organic compounds.
\begin{figure}
	\begin{center}
		\caption{From \cite{Jadzyn2008_02} - Temperature dependence of the static dielectric permittivity measured for C$_6$H$_{13}$PhPhCN and n-hexylcyanobiphenyl mixtures in the isotropic (for the whole concentration range) and the nematic (for $0 \leq x \leq 0.3$) phases.}
		\label{FigJadzynFig4_2008_02}
	\end{center}
\end{figure}
\begin{figure}
	\begin{center}
		\caption{From \cite{Jadzyn2008_02} - Critical-like temperature dependence of the static permittivity derivative ($d\e_s/dT$) and the corresponding field-induced increment of the FE (following Eq. (\ref{s})) for C$_6$H$_{13}$PhPhCN in n-hexylcyanobiphenyl mesogenic solutions in the prenematic region. The derivative is calculated from the plot \ref{FigJadzynFig4_2008_02}.}
		\label{FigJadzynFig6_2008_02}
	\end{center}
\end{figure}

\subsection{Organic molecular crystals}\label{MolCrys}
Several organic semiconductors raised great interest in research, due to high charge mobilities making them promising materials for applications in electronics \cite{deBoer2003,Verlaak2004,Pivovar2006,Podzorov2004}. Among the most studied ones there are rubrene, tetracene, pentacene, and $\alpha$-quaterthiophene, which display polymorphism and several different crystalline phases very close in temperature and energy \cite{Gundlach1999,Siegrist2001,Sondermann1985,DellaValle2006}. This polymorphic behavior requested a great experimental effort to carefully mapping the material response as a function of temperature and thermodynamic history, for which several techniques were used. In this vein, FE was employed to study the phase evolution and correlated features in molecular crystals. Namely, two pioneering works were carried out on films of $\alpha$-quaterthiophene (4T) and tetracene (TEN) crystals \cite{Sassella2011,Parravicini2012}.

\subsubsection{Phase transitions}\label{MC-PT}
Organic crystalline semiconductors (OCS) typically undergo phase transitions depending on temperature variation \cite{Tavazzi2006,Campione2007}. Their evolution displays peculiar features: namely, the onset of phase transformations typically are correlated to occurring of instabilities, due to a gradual increase of molecular disorder, causing macroscopic effects. Such phenomena show analogous features to the pretransitional (or premelting) effects, as e.g. in the previously discussed Ga nanoparticles (section \ref{Ga_nanopart}), and can be found in temperature intervals as wide as tens of degrees straddling the specific transition \cite{Parravicini2003APL}. These pretransitional effects in OCS are usually neglected because they are difficult to be detected. However, taking into account them is mandatory when the critical temperatures and the corresponding instability extension of phases must be identified. Namely, when either low-dimensional or  micro- (and nano-) metric samples are considered (thin films, powders, etc.), similarly to what happens in metallic particles, the features of transitions are expected to be correlated to the involved sizes of the considered system. In this vein, FE evaluation can be fruitfully exploited. Specifically, in $\alpha$-quaterthiophene (4T) sample, FE behavior is able to provide information about the nature and temperature evolution of the structural transitions. The temperature dependence of the permittivity of single crystal $\alpha$-quaterthiophene is shown in Fig. \ref{FigSassella2011}a and the corresponding FE in Fig. \ref{FigSassella2011}b.
\begin{figure}
	\begin{center}
		\caption{From \cite{Sassella2011} - (a) Experimental dielectric constant $\e_s$ and ac conductivity $\sigma$ (in logarithmic scale) as a function of temperature T for a 4T sample. (b) FE contribution $S_E/E^2$ as a function of $T$, obtained from the derivative of the experimental dielectric constant $\e_s$ in the range temperature region 180-225 $^{\circ}$C where the solid-solid and solid-liquid transitions occur; the horizontal lines indicate mean derivative values before and after the two main transitions while the vertical dotted lines mark the transition temperatures.}\label{FigSassella2011}
	\end{center}
\end{figure}

From the Plot \ref{FigSassella2011}a we can see that for the temperature interval between 25 and 190 $^\circ$C, the $\e_s$ dielectric function is a nearly constant, with a slight increase in $T$. At about 190 $^\circ$C, $\e_s$ increases more rapidly with $T$, then, at $T=201$ $^\circ$C, it displays a very sharp enhancement. Above 201 $^\circ$C, the displayed curve further exhibits a positive slope up to $T_{m}=217$ $^\circ$C and, finally, for $T>T_{m}$ the $\e_s(T)$ curve slope becomes negative.
\noindent When FE is considered (Fig. \ref{FigSassella2011}b), a first signal of a transition phenomenon is pointed out by the increasing, starting at 190 $^\circ$C, of $S_E$; then the subsequent very sharp peak at $T=201$ $^\circ$C represents the critical transition temperature. Above $204$ $^\circ$C, a region where $S_E$ still remains positive and constant is found. This demonstrates the occurring of a structural transformation between two distinct ordered-solid stable phases with slightly different degrees of order. We observe that this transition starts at 190 $^\circ$C and it ends at 204 $^\circ$C, with a temperature extension of about 15 $^\circ$C.
\noindent Furthermore, when the temperature arrives to $T_{m}=217$ $^\circ$C, FE displays a step-like discontinuity without significant pretransitional effects, where entropy changes its sign from positive to negative. Congruently with the discussed Fr\"{o}hlich's interpretation, this trend points out a transition from an ordered solid phase to a disordered liquid-like phase. Finally, for $T>T_m$, FE displays negative constant value, which is the signature of a stable liquid-like system. It is worth noting that the induced $S_E$ value calculated for 4T in its liquid phase is comparable to the value already found in other organic liquids where FE has been evaluated \cite{Jadzyn2007}, although it contains different kinds of polar molecules.

Indeed, we can conclude that in OCS the FE evaluation technique is able to detect the nature and evolution of temperature-induced phase transition with its extent in temperature and to give a measure of the state of order of solid and liquid state.

%

\subsubsection{Stability of phases}
The matter of evaluating the stability degree of a given matter phase has been specifically faced on several physical systems. The usual approach is based on thermodynamic arguments such as Gibbs free energy, entropy, enthalpy and analogous potentials \cite{Landau,LandauSTAT,Callen,Kitajgorodskij1965,Miracle2017,Muscat2002,Guo2011,Tessier2000,Sethna2006} which involve several experimental techniques. Congruently, the features of FE evaluation technique make it a good candidate to obtain information about the stability. Actually organic molecular crystals are a paradigmatic class of systems where the stability of phases is an open issue.

A detailed stability analysis was accomplished in TEN crystals \cite{Parravicini2012}, where the coexistence of different phases during large temperature intervals had previously been reported \cite{Sondermann1985}. Moreover, this organic compound can be considered a test system because it was investigated by several different experimental techniques, such as X-ray and neutron diffraction, optical and Raman spectroscopies; they revealed a rich polymorphism in correlation with thermodynamic parameters \cite{Sondermann1985,Pivovar2006,Prikhotko1996,Vaubel1970,Kolendriski1979,Glinski1981,Jankowiak1979,Venuti2004}. The use of FE estimation on TEN, in comparison with the literature, allowed us to clarify several open issues \cite{Parravicini2012}. Experiments are carried out by performing dielectric measurements in a wide temperature region (12-295 K) reported in Figs. \ref{FigParravicini2012-01} and \ref{FigParravicini2012-02} for the cooling and heating stages respectively. In Fig. \ref{FigParravicini2012-01}a (cooling) three peaks of the permittivity at 50, 153, and 223 K are shown that point out the occurring of transitions; a strongly perturbed region is found between 125 and 160 K. The corresponding FE plot (Fig. \ref{FigParravicini2012-01}b) displays a transition signature at the same temperatures. 

Below and immediately above the instability region, the features of three approximately stable phases, with different levels of order, are evidenced in a near flat FE trend. Above 223 K the continuous increasing with temperature of the FE positive values demonstrates a progressive strong increase of the correlations between molecules. In the heating stroke (Fig. \ref{FigParravicini2012-02}) a marked different behavior is evidenced. Inspected peaks of permittivity response (Fig. \ref{FigParravicini2012-02}a) are much higher and sharper and no perturbed temperature region between peaks is evidenced. FE plot, on the other hand, points out five well-defined phase transition signatures separating six phases (Fig. \ref{FigParravicini2012-02}b). The three phases at lowest temperatures (25-45 K, 45-105 K, 105-144 K) display a FE with positive sign and nearly flat trend, indicating a global stability of these phases. Conversely, the highest temperature phases (144-178 K, 178-207 K and 207-280 K) show strongly changing order state. In particular, the highest temperature phase maintains negative FE values, so indicating a strongly-disordered state. The instability region inspected in cooling (Fig. \ref{FigParravicini2012-01}b) is interpreted as an overlapping of the two pointed out phases between 105 K and 178 K \cite{Parravicini2012}.
%
%
%
\begin{figure}
	\begin{center}
		\caption{From \cite{Parravicini2012} - (a) Temperature dependence of the relative dielectric constant $\e_r$ and of the conductivity $\sigma$, as obtained from the dielectric measurements during cooling from RT down to 12 K of a 3-$\mu$m-thick TEN single crystal in the plane parallel capacitor configuration sketched in the inset. (b) Temperature dependence of $s=s(T)$, obtained from the derivative of the dielectric constant in panel (a). We note that the instability in the 125-160 K region is due to the overlapping of two well-distinct transitions which are disclosed in the heating stroke depicted in plot \ref{FigParravicini2012-02}. In both panels the added vertical dotted red lines indicate the critical temperatures 50, 153, and 223 K.}\label{FigParravicini2012-01}
	\end{center}
\end{figure}
\begin{figure}
	\begin{center}
		\caption{From \cite{Parravicini2012} - (a) Temperature dependence of the relative dielectric constant $\e_r$ and of the conductivity $\sigma$, as obtained from the dielectric measurements during heating from 12 K up to RT of the same sample as in Fig. \ref{FigParravicini2012-01} and with the same temperature scan rate. Inset: same as in panel (a), but obtained upon heating the sample at a higher rate. (b) Temperature dependence of $s=s(T)$, obtained from the derivative of the dielectric constant in panel (a). In both panels the added vertical dotted red lines indicate the critical temperatures 45, 105, 144, 178, and 207 K.}\label{FigParravicini2012-02}
	\end{center}
\end{figure}

Finally, we also note that results of Figs.  \ref{FigParravicini2012-01} and \ref{FigParravicini2012-02} reveal a clear non-ergodic response, i.e. a strong dependence on the thermodynamical history of the sample \cite{Parravicini2012,Bokov2006}. This can be found both in the change of critical temperatures of the inspected transitions and in the strong change of the state of order of the phases.

\subsection{Disordered perovskite crystals}
The physical meaningfulness of FE analysis was pinpointed by its application to disordered ferroelectric single crystals, where this experimental approach was also tested to highlight anisotropic features. The analysis was specifically employed for describing the thermal evolution of long-range phases and state of order in inorganic perovskites for photonics applications \cite{Parravicini2016,Tan2019,LoPresti2020,DelRe2015,Parravicini2017,Huang2019,Parravicini2011OE,Wang2020,Parravicini2020,Huang2020}, where peculiar dielectric, optical, thermodynamic, and structural effects are still under investigation \cite{DelRe2015,Parravicini_J2_2012,Parravicini2009,DelRe2011,ParraviciniOL2012,Parravicini2013,Pierangeli2015,Pierangeli2016,DiMei2018,Falsi2020}. These compounds are known to host\footnote{below the so-called \emph{Burns temperature} \cite{Bokov2006,Samara2003,Bokov2012,Toulouse2008,Burns1983}} nanosized reconfigurable polar regions, the so-called \emph{polar nanoregions} (PNRs), which are mesoscopic regions with inherent electric polarization. These can reconfigure in different states as liquid, solid, and even glassy systems \cite{Bokov2006,Samara2003,Pirc2014}. Moreover, due to their high dipole moment, they are able to generate giant dielectric susceptibilities ($\sim 10^4-10^5$). This means that the dielectric response of these materials is essentially due to PNRs and, therefore, that dielectric measurements are the specific technique for investigating their complex and varied evolution, so FE analysis is particularly appropriate \cite{Bokov2006,Bokov2012,Toulouse2008,Kao2004,LoPresti2020}.

\subsubsection{Phase transitions}
Compositionally disordered perovskite crystals display a sequence of structural symmetry transformations as a function of temperature \cite{Jona1993,Lines2001}. Starting from high temperature, the sequence of available configurations\footnote{depending on the specific kind of perovskite} is cubic, tetragonal, orthorhombic, rhombohedral. The paradigm of these systems is Barium Titanate (BaTiO$_3$) \cite{Jona1993,Kao2004,Lines2001}, which displays all these crystalline symmetries. Several works have been devoted to investigate phase transitions in these compounds, e.g. by using NMR and EXAFS techniques \cite{Zalar2003,Zalar2005,Stern2004}. Namely, as hinted above, PNRs play a key role both in the phase stability and in transition processes. Their structural symmetry, which are different from that of the surroundings, was explored in the tetragonal and orthorhombic phases and they were found to be rhombohedral \cite{Tsuda2012,Tsuda2013}. Nonetheless the features of the PNRs in the low-temperature crystalline phases remain, at present, only partially explored for a very restricted number of compounds (e.g. BaTiO$_{3}$, SrTiO$_{3}$, KNbO$_{3}$ and PbTiO$_{3}$) \cite{Bokov2006,Samara2003,Bokov2012,Toulouse2008}. In this vein, the FE evaluation was employed in potassium and tantalum based perovskitic crystals across the cubic-to-tetragonal phase transition \cite{Tan2019,DelRe2015,Huang2019}. Actually, works about KTa$_{1-x}$Nb$_x$O$_3$ (KTN) are focused on this transition \cite{Tan2019,Huang2019,Huang2020}. Fig. \ref{FigTan2019} from \cite{Tan2019} points out that when the system passes to a paraelectric state FE evolution signals an order/disorder transition. This demonstrates that a crystalline system following the Curie-Weiss law can be dielectrically considered a liquid of dipoles.
\begin{figure}
	\begin{center}
		\caption{From \cite{Tan2019} - (Top) Dielectric response of KTN and (bottom) correlated FE (following Eq. (\ref{s})) across the ferroelectric-to-paraelectric transition. Note that FE plot clearly evidences the critical $T_C$.}\label{FigTan2019}
	\end{center}
\end{figure}
These concepts are developed in a much wider temperature region in papers about K$_{1-y}$Li$_y$Ta$_{1-x}$Nb$_x$O$_3$ (KLTN) \cite{Parravicini2016,DelRe2015,Parravicini2020} and K$_{1-y}$Na$_y$Ta$_{1-x}$Nb$_x$O$_3$ \cite{Parravicini2017}. The considered samples display four crystalline phases and three transitions along the 320-25 K range, whose dielectric response measurement (Fig. \ref{Fig2Parravicini2016} for KLTN) and FE evaluation (Fig. \ref{Fig3Parravicini2016} for KLTN and Figs. \ref{FigHuang2020}c, \ref{FigHuang2020}d for KTN) allows to continuously follow the temperature evolution. This treatise point out a complex mixture of ordered and disordered phases. We note that FE evaluation enables to identifies all critical temperatures with high precision and points out that transitions are associated to order-disorder changes. More specifically, in KLTN, above each transition, some disordered regions are evidenced (Fig. \ref{Fig3Parravicini2016}). This phenomenon can be correlated to the arrangement of PNRs, which in this region are free-to-reorient, being in a liquid-like state. When, in correspondence to the order-disorder change (in the positive FE sign) they become correlated, they produce a structured (ordered) configuration that some authors indicate as a glassy state \cite{Toulouse2008,Pirc2014,Cai2015}. This PNRs configuration continuously evolves and below the FE peak they form large polarized domains giving rise to the ferroelectric arrangement of the tetragonal phase (paraelectric to ferroelectric transition) \cite{Fu2012}.

Finally, we note that FE oscillations can also be found in such a kind of physical systems (Fig. \ref{FigDelRe2015NP}), whose physical meaning is the same as in the previously discussed molecular crystals (section \ref{MolCrys}); they point out specific instability regions where, e.g., non-ergodic dynamics can arise (Fig. \ref{FigParravicini2012-01}) \cite{Parravicini2012,Parravicini2020}.
\begin{figure}
	\begin{center}
		\caption{From \cite{Parravicini2016} - KLTN relative dielectric permittivity as a function of temperature $T$ measured at $10^3$-$10^6$ Hz frequencies along $a$ (a), $b$ (b), and $c$ (c) perpendicular directions for the cooling stage.}\label{Fig2Parravicini2016}
	\end{center}
\end{figure}
\begin{figure}
	\begin{center}
		\caption{From \cite{Parravicini2016} - KLTN FE (following Eqs. (\ref{s}) and (\ref{s_direzionale})) $s_a$, $s_b$, $s_c$, calculated by applying Eq. (\ref{s}) to the data displayed in Fig. \ref{Fig2Parravicini2016}, for electric fields applied along, respectively, $a$ (a), $b$ (b), $c$ (c) axes directions (corresponding to $[100]$, $[010]$, and $[001]$ crystalline directions). The indicated temperatures $T_C^{'}$, $T_C^{''}$, $T_C^{'''}$ are the identified transition temperatures. The insets show the temperature range 0-100 K for each respective direction and they point out that the anisotropy tend to vanish as the temperature approaches 0 K.}\label{Fig3Parravicini2016}
	\end{center}
\end{figure}
\begin{figure}
	\begin{center}
		\caption{From \cite{Huang2020} - (a, b) The curves of $\e_r$ versus temperature for two samples of potassium tantalate-niobate (KTN1 and KTN2) which were grown in different strain conditions (see reference). (c, d) The Fr\"{o}hlich entropy $s$ for samples cut along different directions from KTN1 and KTN2, respectively, the black dotted lines represent $s=0$. The dielectric permittivity was measured at 1 kHz.}\label{FigHuang2020}
	\end{center}
\end{figure}

\newpage
\subsubsection{Evolution of the directional order state}
The description of the order state evolution depending on $T$ was proven particularly valid in describing anisotropy; this approach can be considered quite original. Actually, although the foundation theory is mathematically easy and well established \cite{Devonshire1949,Scott2011,Nye1985,Devonshire1954,Landau}, nevertheless it is infrequent to consider thermodynamic quantities as a function of directional physical variables. However, works about disordered perovskites consider FE behavior as a function of crystalline directions, to highlight anisotropic features of their order/disorder state \cite{Parravicini2016,Parravicini2017,Huang2019,Huang2020}. Actually, these studies exploit the theory of thermodynamic potentials in anisotropic dielectrics which was systematically developed by J. Parravicini \cite{Parravicini2018}. Namely, we can associate the directional FE evaluation therein with Eqs. (\ref{Rel}) and (\ref{s_direzionale}). The works on KLTN \cite{Parravicini2016}, KNTN \cite{Parravicini2017}, and KTN \cite{Huang2020} point out that these systems undergoes several order/disorder transitions and, in particular, that the feature of either order or disorder depends both on the crystalline phase and the considered direction along which the electric probing field is applied \cite{Parravicini2016,Parravicini2017,Huang2020}. Actually, these materials have temperatures where they behave as ordered systems in one direction and disordered systems in another direction (Fig. \ref{Fig3Parravicini2016}). This is found in tetragonal and orthorhombic phases, which display unusual arrangement where the system is contemporaneously solid-like in one direction and liquid-like in another. Congruently, these investigations point out that just the highest (nominally cubic) and lowest (nominally rhombohedral) crystalline phases have the same order state in all directions, i.e. disordered and ordered feature respectively (Fig. \ref{Fig3Parravicini2016}). Being the dielectric response of such systems essentially due to the PNRs, in \cite{Parravicini2017} the authors infer that the correlated FE identify an exotic arrangement of them, whose schematic depiction is reported in Fig. \ref{FigParravicini2017}.
\begin{figure}
	\begin{center}
		\caption{From \cite{Parravicini2017} - Schematic representation of the nanometric dipole arrangement in the four nominal phases of KNTN (it holds for KLTN also). On the opposite ends the two standard fully solid or fully liquid states: the rhombohedral phase with three ordered (solid) directions (ordered, ordered, ordered, ``ooo''): global view (d1) and projections along the axes $x$ (ordered, solid-like, a1), $y$ (ordered, solid-like, b1) and $z$ (ordered, solid-like, c1); the cubic phase with three disordered (liquid) directions (disordered, disordered, disordered, ``ddd''): global view (d4) and projections along the axes $x$ (disordered, liquid-like, a4), $y$ (disordered, liquid-like, b4) and $z$ (disordered, liquid-like, c4). In the middle the two exotic liquid-solid directional composite states: the orthorhombic phase with two ordered (solid) and one disordered (liquid) directions (disordered, ordered, ordered, ``doo''): global view (d2) and projections along the axes $x$ (disordered, liquid-like, a2), $y$ (ordered, solid-like, b2) and $z$ (ordered, solid-like, c2); the tetragonal phase with one ordered (solid) and two disordered (liquid) directions (disordered, disordered, ordered, ``ddo''): global view (d3) and projections along the axes $x$ (disordered, liquid-like, a3), $y$ (disordered, liquid-like, b3) and $z$ (ordered, solid-like, c3).}
		\label{FigParravicini2017}
	\end{center}
\end{figure}

It is worth observing that directional FE demonstrates that the rhombohedral phase, which is ordered along all directions, is the true stable phase in these relaxors \cite{Parravicini2016,Parravicini2017,Huang2020}, congruently with what suggested in previous study on similar perovskites \cite{Tsuda2012,Tsuda2013}.

We finally stress that both in KLTN and KNTN a remarkable agreement is found between the distribution of the macroscopic order/disorder state as given by FE with the microscopic arrangement of ions as depicted by the eight-sites model of Com\`{e}s et al. \cite{Comes1968,Comes1970,Parravicini2016,Parravicini2017}.

\subsection{Other systems}
\subsubsection{Polymeric films}
Fr\"{o}hlich's interpretation was also applied in explaining the dielectric relaxation of a polymer-based ferroelectric material, in a work of V.V. Kochervinskii et al. about a copolymer of vinylidene fluoride and tetrafluoroethylene \cite{Kochervinskii2016}. The paper reports the study on this compound in form of textured films, whose it discusses the molecular mobility and structuring process. In particular, the relationship between the orientational polarization and the topology of structure formation is investigated. The relationship (\ref{S_E}) is applied to the dielectric constant at low-frequency. The positive value of $S_E$ in the investigated temperature region evidences a long-range order in the arrangement of kinetic units participating in the dielectric relaxation process of the polymer. These results are in agreement with ferroelectric polymers and with several compounds of this class \cite{Kochervinskii2002_1,Kochervinskii2002_2,Kochervinskii2007}.

\subsubsection{Enzymes}
A further physical system where FE evaluation has been demonstrated to be able to provide useful information are enzymes, namely globular proteins, which are investigated by Kurzweil-Segev et al. in the framework of complex systems \cite{Kurzweil2016}. Actually, the work treats hydrated lysozyme powders with different amount of confined water. Authors carry out a comparison between dielectric and calorimetric measurements in a broad temperature region. The FE data are shown in Fig. \ref{FigKurzweil2016} and the correlated differential scanning calorimetry (DSC) measurements are depicted in Fig. \ref{FigKurzweil2016Fig4}. Comparison of the results from two samples with different water confinement levels is depicted. By correlating dielectric and calorimetric data, the authors remarkably associate the sign change of FE, from negative to positive values, to a glass transition, indicating that complex cooperative processes are acting in hydrated enzymes \cite{Kurzweil2016}.
\begin{figure}
	\begin{center}
		\caption{From \cite{Kurzweil2016} - FE evaluation for two samples (gray circles and black squares, respectively). The parallel tendency of the dipole orientations takes place in the range within the blue background and the short-range orientation is found within the yellow background.}\label{FigKurzweil2016}
	\end{center}
\end{figure}
\begin{figure}
	\begin{center}
		\caption{From \cite{Kurzweil2016} - DSC cooling thermograms of the enzymes sample with (black curve) and without (grey curve) annealing.}\label{FigKurzweil2016Fig4}
	\end{center}
\end{figure}

\section{Conclusions}
The present review provided the landscape about the exploitation of FE evaluation to inspect the state of order in condensed matter. We have highlighted that this approach, which we call \emph{Fr\"{o}hlich entropy technique}, can be considered a reliable experimental approach for the investigation of several physical systems. Summarizing, the presented method allows the following items:
\begin{enumerate}%
	\item observation of the thermal evolution of the state of order in the sample almost continuously;%
	\item determination of a stability level of the explored phase;%
	\item very punctual detection of the phase transitions, critical temperatures with correlated neighbourhoods and ordering features;%
	\item evaluation criterion about the correlation state and consequent rotational freedom of the inherent dipoles;
	\item directional investigation highlighting possible anisotropies;%
	\item relatively easy and not expensive application for the investigation of a wide class of different physical systems in condensed matter.
\end{enumerate}
The presented \emph{Fr\"{o}hlich entropy estimation technique}, which allows a virtually continuous detection in temperature of the explored sample, on one hand has macroscopic nature, on the other hand gives the possibility to reconstruct information both about macroscopic and mesoscopic phenomena. Then, this technique appears a powerful method complementary to the largely exploited approaches based on local investigation at a fixed temperature.

\section*{Acknowledgements}
We acknowledge Prof. A. Stella for his contributions in conceiving the FE evaluation technique. %
We finally thank V. Sanvito and M. Bartozzi for their constant support.

\bibliographystyle{elsarticle-num}
\bibliography{mybibfile}






\end{document}